\pdfoutput=1
\documentclass[conference]{IEEEtran}
\IEEEoverridecommandlockouts
\usepackage{amsmath,amssymb,amsfonts}
\usepackage{algorithmic}
\usepackage{graphicx}
\usepackage{textcomp}
\usepackage{xcolor}
\usepackage[numbers]{natbib}
\usepackage{listings}
\usepackage{inconsolata}
\usepackage{comment}
\usepackage{amsfonts} 
\usepackage{textcomp}
\usepackage[utf8]{inputenc}
\usepackage[free-standing-units,per-mode=symbol,mode=text,reset-text-series=false,reset-text-family=false,text-series-to-math=true,text-family-to-math=true]{siunitx}
\usepackage{tikz} 
\usepackage{glossaries}
\usepackage{import} 
\usepackage{microtype}
\usepackage{url}
\usepackage[inline]{enumitem}
\usepackage{multirow}
\usepackage{booktabs}
\usepackage[capitalise,noabbrev,nameinlink]{cleveref}
\usepackage{subcaption}
\usepackage[frozencache,cachedir=minted-cache]{minted}\usepackage[export]{adjustbox}
\usepackage{hyphenat}
\usepackage{svg}

\definecolor{keywordcolor}{rgb}{0.1, 0.6, 0.1}  
\definecolor{commentcolor}{rgb}{0.5, 0.5, 0.5}  
\definecolor{stringcolor}{rgb}{0.6, 0.1, 0.1}   
\definecolor{backgroundcolor}{rgb}{0.97, 0.97, 0.97} 

\lstdefinestyle{customstyle}{
    backgroundcolor=\color{white},
    commentstyle=\color{MidnightBlue},
    keywordstyle=\color{keywordcolor}\bfseries,
    stringstyle=\color{stringcolor},
    basicstyle=\ttfamily\footnotesize,
    morekeywords={for, in, to}, 
    breakatwhitespace=false,         
    breaklines=true,                 
    captionpos=b,                    
    keepspaces=true,                 
    numbersep=5pt,                  
    showspaces=false,                
    showstringspaces=false,
    showtabs=false,                  
    tabsize=2,
    escapeinside={(*@}{@*)}, 
}

\def\BibTeX{{\rm B\kern-.05em{\sc i\kern-.025em b}\kern-.08em
    T\kern-.1667em\lower.7ex\hbox{E}\kern-.125emX}}
\begin{document}

\title{SpikeStream: Accelerating Spiking Neural Network Inference  on RISC-V Clusters with Sparse Computation Extensions\\}

\author{
    \IEEEauthorblockN{Simone Manoni\IEEEauthorrefmark{1}, Paul Scheffler\IEEEauthorrefmark{2}, 
    Luca Zanatta\IEEEauthorrefmark{3}, 
    Andrea Acquaviva\IEEEauthorrefmark{1},
    Luca Benini\IEEEauthorrefmark{1}\IEEEauthorrefmark{2},
    Andrea Bartolini\IEEEauthorrefmark{1}}
    \IEEEauthorblockA{\IEEEauthorrefmark{1}\textit{Department of Electrical, Electronic and Information Engineering, University of Bologna}}
    \IEEEauthorblockA{\IEEEauthorrefmark{2}\textit{Integrated Systems Laboratory, ETH Zürich}}
    \IEEEauthorblockA{\IEEEauthorrefmark{3}\textit{Autonomous Robots Lab, Norwegian University of Science and Technology (NTNU)}}
    \IEEEauthorblockA{\{s.manoni, andrea.acquaviva, a.bartolini\}@unibo.it, \{paulsc, lbenini\}@iis.ee.ethz.ch, luca.zanatta@ntnu.no}
}

\maketitle

\begin{abstract}
Spiking Neural Network (SNN) inference has a clear potential for high energy efficiency as computation is triggered by events. However, the inherent sparsity of events poses challenges for conventional computing systems, driving the development of specialized neuromorphic processors, which come with high silicon area costs and lack the flexibility needed for running other computational kernels, limiting widespread adoption.
In this paper, we explore the low-level software design, parallelization, and acceleration of SNNs on general-purpose multicore clusters with a low-overhead RISC-V ISA extension for streaming sparse computations. We propose SpikeStream, an optimization technique that maps weights accesses to affine and indirect register-mapped memory streams to enhance performance, utilization, and efficiency. 
Our results on the end-to-end Spiking-VGG11 model demonstrate a significant 4.39× speedup and an increase in utilization from 9.28\% to 52.3\% compared to a non-streaming parallel baseline. Additionally, we achieve an energy efficiency gain of 3.46$\times$ over LSMCore and a performance gain of 2.38$\times$ over Loihi.
\end{abstract}





\section{Introduction}
Spiking Neural Networks (SNNs) are at the forefront of neuromorphic computing, providing compact and energy-efficient AI models \cite{neurorv}.
Recent studies have demonstrated that SNNs are capable of achieving competitive performance in image classification and object detection tasks with a reduced power envelope and memory footprint w.r.t. their non-spiking counterparts \cite{sobjdetect, ptb, cdnn}. 
To address their event-driven nature, spike-based communication between neurons, and complex activation functions, several accelerators and neuromorphic processors have been developed. Among these, both Von Neumann architectures and non-Von Neumann architectures have been proposed.
The latter are composed of neurons and synapses arrays \cite{neuroopp}, and can be grouped into analog and mixed-signal processors \cite{dynap, neurogrid}, Globally Asynchronous and Locally Synchronous (GALS) processors \cite{loihi, odin}, and more traditional Digital Fully-Synchronous (DFS) accelerators \cite{ptb, sobjdetect, cdnn, sne, lsm}. 
Analog neuromorphic processors struggle with precision and noise issues as their continuous signals are vulnerable to voltage and temperature variations \cite{ottati}, leading to less accurate computations compared to digital platforms. GALS processors are very complex to design due to the low maturity of asynchronous EDA tools \cite{neurorv}. 
Both types of processors also face barriers to integration with existing infrastructure, often requiring entirely new software stacks \cite{neurorv}. 
DFS accelerators face fewer integration issues but are often constrained by the types of SNN models they can support, limited by their network topology or their reliance on hardwired neuron models, which frequently use highly-reduced and fixed arithmetic precision \cite{sne, lsm, cdnn}. 

Moreover, quantization techniques in SNNs are not yet as stable as their ANNs counterparts and can lead to significant accuracy degradation \cite{zanatta, sobjdetect}. Indeed, while achieving marginal accuracy reduction in simple tasks such as MNIST classification \cite{odin}, Zanatta et al. \cite{zanatta} demonstrate that in a drone obstacle avoidance task, the SNN outperforms the ANN when using floating-point (FP) computation. When the SNN is quantized from FP32 to 4-bit integers, it is no longer able to complete the task. Therefore, FP computation is still beneficial in SNNs. 

Additionally, the rapid evolution of the SNNs ecosystem requires the embrace of programmable and general-purpose (GP) solutions to support flexible model exploration. However, traditional CPUs, face challenges with the event-driven nature of SNNs, resulting in sparse data structures \cite{neurorv} that cause irregular memory access patterns and poor utilization of processing elements \cite{sssr}.

Recently, \textit{Stream Registers} (SRs) \cite{vectstream, ssr, sssr} have emerged as a hardware extension for CPUs to address the Von Neumann bottleneck and the associated inefficiencies with memory accesses. SRs map streams of memory access directly to reads or writes of architectural registers, with address generation and data movement handled by dedicated hardware.
%
%
They maximize bandwidth utilization on memory-bound workloads 
by decoupling memory accesses from computation and continuously streaming useful data, freeing the host processor from address calculations and enabling high floating-point-unit (FPU) utilization and compute throughput. %
In addition to affine address patterns, SRs can also support \emph{indirect} streams \cite{vectstream, sssr} using a base address and an index array to gather or scatter data, which can significantly accelerate the irregular memory accesses of sparse workloads.%
%

In this paper, we present SpikeStream, the first exploration of neuromorphic processing acceleration on a 
multi-core streaming architecture. In contrast to the current State-of-the-Art (SoA) that requires expensive and non-flexible hardware units, our approach is software-based and runs on programmable RISC-V processors with enhanced Instruction Set Architecture (ISA). 
It leverages the low-overhead streaming, SIMD, and hardware-loop extensions of an RV32G parallel compute cluster~\cite{snitch}, to accelerate SNN computation and maximize FPU utilization. 
We first implement a parallel FP SIMD baseline by adopting a compressed representation for the sparse input feature maps (ifmaps). We identify the \textit{indirection} operation used to gather weights associated with input spikes as the main source of inefficiency in neuromorphic processing on CPUs, leading to frequent address computations, irregular memory accesses, and loop control overhead.
SpikeStream optimizes SNN computation by leveraging the available \cite{sssr} extensions in the proposed architecture to address memory inefficiencies caused by the indirection operation. When compared with the parallel baseline implementation of the SNN computation, SpikeStream improves runtime inference on an S-VGG11 of 7.29$\times$ and introduces an energy-efficiency gain of 5.68$\times$ in FP8, increasing the FPU utilization from 9.28\% to 52.3\%.

When compared with neuromorphic processors, SpikeStream achieves competitive performances, outperforming Loihi by 1.31$\times$ in FP16 and 2.38$\times$ in FP8, and introducing an energy-efficiency gain over LSMCore of 2.37$\times$ in FP16 and 3.46$\times$ in FP8, demonstrating that our approach can significantly improve the performance, utilization, and energy-efficiency of neuromorphic processing even without ad-hoc hardware. 

\section{Background}
\subsection{Spiking Neural Networks}

SNNs employ spiking neurons as their fundamental computational units. The Leaky Integrate-and-Fire (LIF) model is a widely adopted model of a spiking neuron since its simplicity is appealing in deep learning tasks.
The following equation governs the LIF model:
\begin{equation}
    \begin{cases}
        i_m(t) = \sum_{n=1}^N s_{i, n}(t) w_n \\
        v_m(t) = v_m(t-1)\alpha + r i_m(t) - v_{rst} s_{o, m}(t) \\
        s_{o, m}(t) = \begin{cases}
            1 & \text{if } v_m(t) \ge v_{th} \\
            0 & \text{otherwise}
        \end{cases}
    \end{cases}
\end{equation}
where, \( s_{i,n}(t) \) and \( s_{o,m}(t) \) represent the input spike at synapse \( n \) and the output spike of the \( m \)-th neuron at time step \( t \), respectively. \( w_n \) is the weight associated with synapse \( n \), and \( N \) is the total number of synapses. \( i_m(t) \) and \( v_m(t) \) denote the input current and neuron state of the output neuron \( m \), respectively. \( v_r^{st} \), \( v_{th} \), \(r\) (usually set to 1), and \( \alpha \) are the reset potential, membrane threshold, membrane resistance, and decay factor, respectively.

\subsection{The multi-core streaming architecture} \label{back2}
The adopted multi-core streaming architecture is the open-source \emph{Snitch} cluster~\cite{snitch}, a programmable many-core accelerator targeting energy-efficient FP compute.
It contains eight 32-bit RV32G \emph{worker} cores, each featuring a SIMD-capable FP64 FPU kept busy by three SRs and a hardware loop that decouples the FPU and integer core. These two subsystems are synchronized by explicit move instructions, allowing Snitch to overlap independent integer and FP instructions. SRs also expose shadow registers to overlap configuration and computation. 
The cluster also provides a shared 128 KiB 32-bank scratchpad memory (SPM), accessible to worker cores through a single-cycle logarithmic interconnect, and a shared 8 KiB L1 instruction cache.
An additional \emph{DMA} core without SRs and FPU controls a 512-bit DMA engine used to asynchronously move large tiles of data between the cluster's SPM and global memory.

The SRs map buffered streams to and from the cluster's SPM~\cite{ssr} directly to FP register reads and writes, handling all necessary address calculations in hardware.
All three SRs in each worker core are capable of $\leq$4D affine streams, enabling near-constant FPU utilization on dense workloads exhibiting regular memory access patterns.
Furthermore, two of them support 1D indirect streams with 8-, 16-, or 32-bit indices in SPM~\cite{sssr}, which can significantly accelerate workloads involving sparse and irregular data structures. 
Refer to \cite{sssr} for more details on Snitch Cluster architecture.
\section{SpikeStream Software Architecture} \label{sec:implementation}
  This section presents the proposed optimized SpikeStream SNN inference kernel leveraging the architecture presented in section \ref{back2}. We progressively describe the key optimizations implemented: Tensor compression (TC - Section \ref{TC});  Task parallelization (TP - Section \ref{TP}); Data parallelization (DP - Section \ref{DP}); Tiling and double buffering (DB - Section \ref{DB}) and streaming acceleration (SA - Section \ref{SA}). 
\renewcommand\fcolorbox[4][]{\textcolor{cyan}{\strut#4}}
\begin{figure}[ht!]
\hspace{1em}
\begin{minipage}{0.4\textwidth} 
%
\begin{minipage}[t]{\linewidth}
\begin{minted}[fontsize=\fontsize{6.6}{7.0}\selectfont]{python}
# core_rf_start, w_baddr set by workload-stealing scheduler
# Spatial iterations on RF
for i in range(k * k):
    # Select ifmap row
    if i % k == 0 and i != 0:
        if_rptr += 1
    # Compute spatial coordinate for s_ptr_i
    coo_ptr = if_rptr * if_w + core_rf_start + (i % k)
    # Compute stream/SpVA base address
    s_baddr = s_ptr_i[coo + 1]
    # Compute stream length
    s_len = s_baddr - s_ptr_i[coo]
    # Execute SpVA
    for j in range(s_len):
        # Accumulate on output neuron's input current
        ic += w[c_idcs_i[s_baddr + j] + w_baddr]
# Apply activation function
act_fun(ic, v, vth, c_idcs_o, s_ptr_o)
\end{minted}
\subcaption{Pseudocode of the spatial iterations on the RF.}
\label{lst:base}
\end{minipage}
%
\begin{minipage}[t]{\linewidth}
\vspace{0.45em}
\begin{minted}[fontsize=\fontsize{6.6}{6.8}\selectfont]{gas}
SpVA: lw     t0, 0(%c_idcs_i)
      slli   t0, t0, 3
      add    t0, t0, %w
      fld    ft1, 0(t0)
      addi   %c_idcs_i, %c_idcs_i, 2
      addi   %iter, %iter, 1
      fadd   %ic, ft1, %ic
      bne    %iter, %s_len, SpVA
\end{minted}
\subcaption{SpVA loop RISC-V assembly.}
\label{lst:spvarisc}
\end{minipage}

\begin{minipage}[t]{\linewidth}
\vspace{0.45em}
\begin{minted}[fontsize=\fontsize{6.6}{6.8}\selectfont]{gas}
if s_len != 0:
        sr_set_indir(SR1, &w[w_baddr])
        sr_set_idcs(SR1, &c_idcs[s_baddr])
        sr_set_bound(SR1, s_len)
        frep 1, %s_len
        ic += sr_read(SR1)
\end{minted}
\subcaption{SpikeStream SpVA pseudocode}
\label{lst:spikestream}
\end{minipage}
\vspace{-0.3em}
\captionof{listing}{Baseline pseudocode of the spatial iterations on the RF and SpVA loop RISC-V assembly.}
\vspace{-1.4em}
\label{lst:combined_impl}
\end{minipage}
\end{figure}
\subsection{Tensors compression} \label{TC}
In this manuscript, we propose to adopt a memory-efficient fiber-tree-based compression format for ifmaps, derived by the Compressed Sparse Row (CSR) data representation. Indeed, we do not need to store the values of non-zero elements as they are all "1".  In convolutional layers, binary channels are compactly represented using an index array (\texttt{c\_idcs}) to mark the positions of active neurons, while a spatial pointer array (\texttt{s\_ptr}) aggregates information on the spiking neuron count across spatial dimensions. In fully connected (FC) layers, compression is achieved with a single index array and a count of spiking neurons. Furthermore, by processing ifmaps sequentially, we avoid timestamping each input spike, leading to a more compact data representation compared to the address-event-representation (AER) format used in neuromorphic processors, which requires absolute coordinates and a timestamp for each spike \cite{sne}. 
This compressed format allows the replacement of costly multiply-accumulate operations with less power-hungry add operations. 
The neuron state tensor is held in a dense representation as each value is updated at each timestep, resulting in a low degree of sparsity to motivate compression.
In this work, we assume that all input tensors, 
including neuron states, weights and ifmaps 
are stored at the higher level of the memory hierarchy.
\subsection{Task Parallelization} \label{TP}
We parallelized the computational kernel to match the number of working cores in a Snitch Cluster. In the convolutional layers, each worker core is assigned a different receptive field (RF), which is processed in a depth-first manner. To evaluate $i_m(t)$, the kernel initially reads $v_m(t)$ into an FPU register and then computes a Sparse-Dense Vector Accumulation (SpVA) for each spatial iteration in the RF. For each SpVA, the weights associated with an input spike are retrieved through a sequence of indirection operations over the index vectors and accumulated in the register holding $v_m(t)$. The computation of each $i_m(t)$ involves $k_h \times k_w$ SpVAs, with $k_h$ and $k_w$ representing the spatial dimensions of the filters. The number of indirection operations required for each SpVA equals the number of spiking neurons across the channels in each spatial dimension. To address the workload imbalance resulting by using a compressed data representation of ifmaps, we introduced a workload stealing mechanism, in which each core, upon completion of its assigned RF, moves without halting to the next available unprocessed RF via an atomic tagging operation.
Finally, to increase data reuse and reduce transfers to and from external memory, we implement layer fusion with the activation function.
Listing \ref{lst:base} illustrates a streamlined pseudocode of the proposed baseline compressed convolution for SNNs. The two outer loops manage the control-heavy computations required to generate the base addresses for both the weights and \texttt{c\_idcs} which are then accessed by the innermost loop responsible for executing the SpVA. Without any optimizations, the SpVA yields to the assembly code in Listing \ref{lst:spvarisc}, where only 1 instruction does useful computation, while the other 7 instructions are required for address calculations, memory accesses and loop control. 
\subsection{Data parallelization} \label{DP}
Depending on the chosen precision, the SIMD capabilities of the FPU are leveraged to parallelize the activations across the output channels on each core's FPU lanes.
To achieve this, we adopt a batched \textit{HWC} memory layout for the weight tensor, arranging the weights from different filters in contiguous memory locations, with the number of batched weights corresponding to the SIMD width.
Since the activations are parallelized across different FPU lanes, after thresholding, a series of SIMD-width bit-masking and branching operations are required to extract the output neuron result. In the case of a spike, the ofmap's \texttt{c\_icds} and \texttt{s\_ptr} SPM buffers are atomically updated. 

\begin{figure}
    \centering
    \begin{subfigure}{0.48\textwidth}
    \includegraphics[bb=0 0 916 235, width=\textwidth]{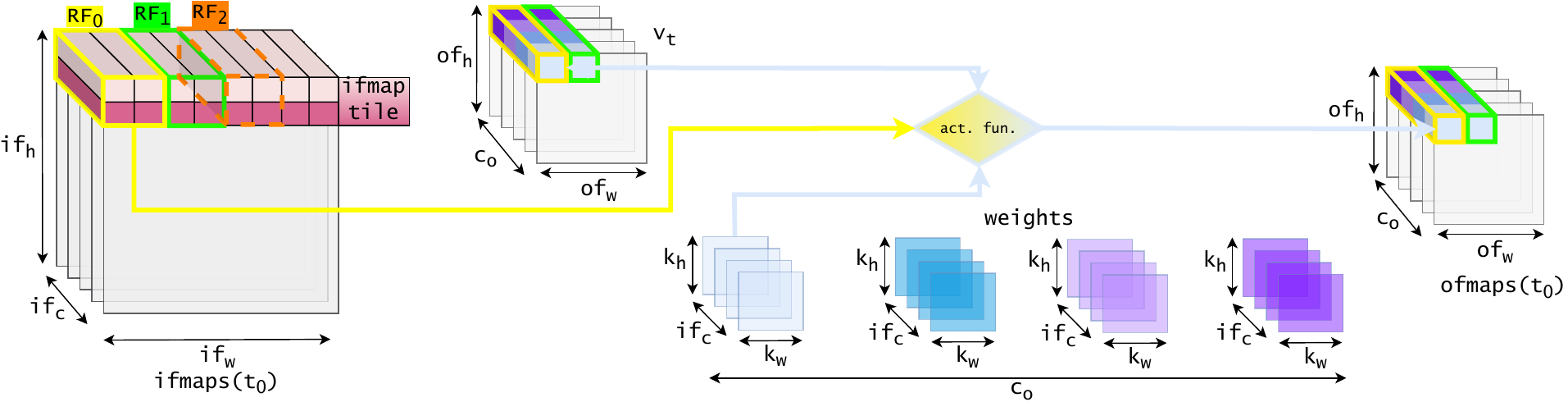}
    \caption{SNN convolutional layer at timestep $t_0$.}
    \label{fig:snn}
\end{subfigure}
    \begin{subfigure}{0.48\textwidth}
    \includegraphics[bb = 26 15 944 375, width=\textwidth]{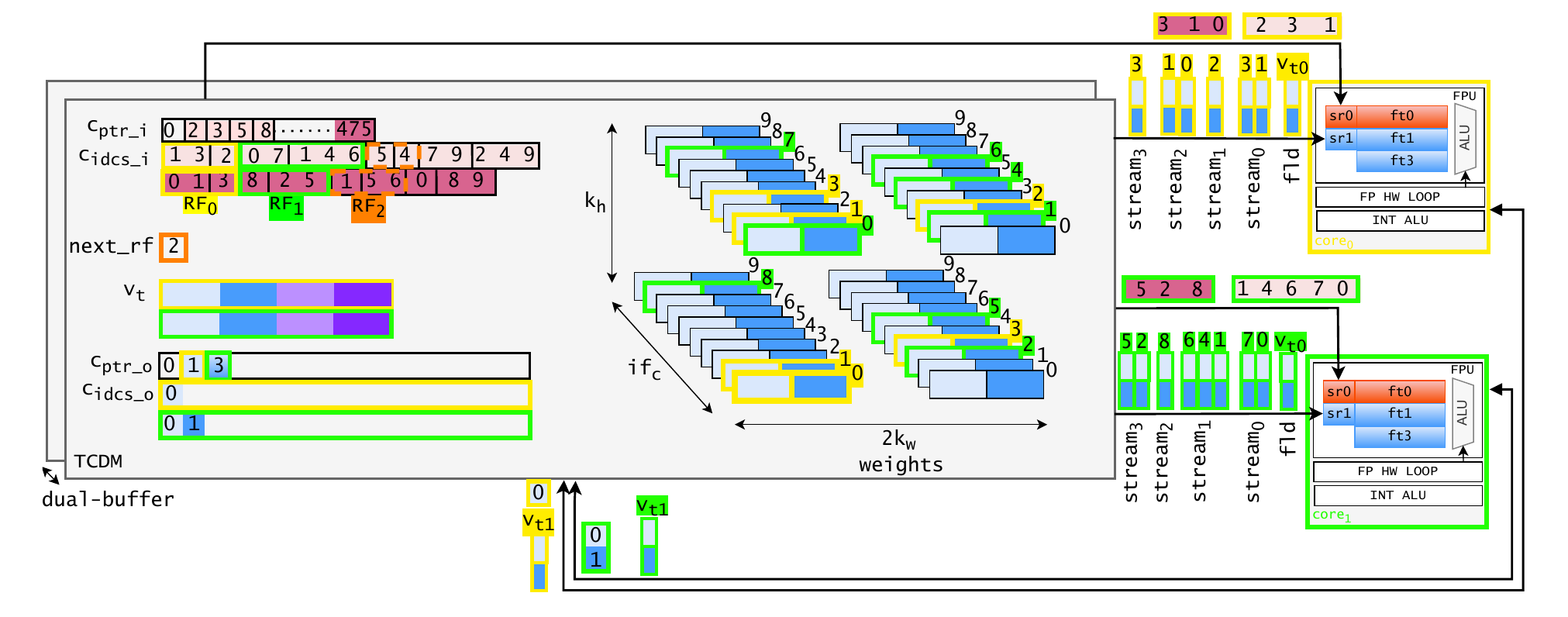}
    \caption{SpikeStream dataflow assuming $k_h=k_w=stride=2$, $2$ \textit{worker cores} and FP32 weights.}
    \label{fig:ssdataf}
\end{subfigure}
\caption{SpikeStream convolutional layer dataflow.}
    \vspace{-1.4em}
\end{figure}

Figure \ref{fig:snn} shows an SNN convolutional at timestep $t_0$ layer with uncompressed tensors. Figure \ref{fig:ssdataf} shows the corresponding SpikeStream DP and TP dataflow using compressed ifmaps and ofmaps. The \texttt{next\_rf} index is used to keep track of the RFs processed and to implement workload stealing scheme.
\subsection{Tiling and double buffering} \label{DB}
We implement double buffering in all kernels to mask the latency associated with memory transfers and maximize utilization.
The kernel process begins with an initialization phase where a tile for each input tensor is loaded into the cluster's SPM. In addition, according to the sizes of the input tiles, SPM buffers are allocated to host the compressed ofmaps computed by the cores. These buffers are sized for the worst-case scenario, assuming a zero-sparsity output. Once this initialization phase is complete, kernel computation begins. Simultaneously, the DMA core preloads the next data chunk from external memory into the SPM in preparation for the next computation phase. 
To avoid excessive fragmentation of the output \texttt{c\_idcs} vectors and to reduce data movement associated with spatial pointers, we first double-buffer the weights and then the ifmaps. This approach ensures that the compressed ofmap tile is fully populated before it is transferred back to external memory.
Once all the weights have been processed for the ifmap tile, the DMA core joins the ofmap \texttt{s\_ptr} elements before copying out the results.
Furthermore, the adopted compression format, which aggregates spiking neuron information into spatial dimensions, allows a compressed ifmap tile to be transferred within a single DMA request. The ofmap \texttt{c\_idcs} vectors still have to be transferred upward individually due to the fragmented output buffer resulting from worst-case allocation caused by dynamic sparsity.
\subsection{Streaming acceleration} \label{SA}
We accelerate neuromorphic processing by reducing the overhead of repeated indirection operations within the SpVA loop, using the SRs to map all indirect weight loads to indexed stream reads. At the start of each SpVA, we configure an indirect SR to point to the base address of the ifmap \texttt{c\_idcs} vector in the dedicated SPM buffer, corresponding to a spatial location within the RF, and the base address of the associated grouped weight tensor. We use \texttt{s\_ptr} to compute the loop trip count to set the stream boundaries and configure the FP repetition buffer on an add operation to ensure continuous streaming and accumulation of 
the weights. This allows SR's hardware units to completely manage the SpVA's address generation and memory accesses. Meanwhile, the hardware-loop unit handles loop control and decouples the FPU from the integer core, allowing it to advance control processing outside of the SpVA to set up the next stream iteration via SR's shadow registers. Listing \ref{lst:spikestream} reports SpikeStream SpVA pseudocode. 
\subsection{Spike encoding}
When working with RGB images rather than event-cameras, it's necessary to convert the images into spikes. Most SNNs achieve this by using the first convolutional layer to handle the conversion, where the raw image values are interpreted directly as input currents of the layer \cite{bellec2020, xiao202, neurips}.
In this case, we prefer a dense representation format with \textit{HWC} storage for the first input tensor.
In SpikeStream, we reshape this tensor on the fly through a 2D DMA transfer, reorganizing it with an \textit{im2row} transformation and converting the convolution into a matrix multiplication (matmul). This allows for easier parallelization across cluster cores by output channel. Each dot product associated with an RF is accelerated using two affine SRs: One for the input current and the other for the weights.
\section{evaluation}
We synthesized the Snitch cluster architecture in GlobalFoundries' 12LP+ FinFET technology using Fusion Compiler 2022.03. Runtime measurements are based on cycle-accurate simulations of the Snitch cluster's register-transfer-level (RTL) description with Questasim 2022.3. All the code tested in the paper are compiled with a customized LLVM 12 toolchain for Snitch, with -O3 optimizations. Simulation traces are then used to extract runtimes and utilization metrics. Energy estimations are obtained by executing the computational kernels in a post-layout gate-level simulation at the cluster's clock speed of 1 GHz. The obtained switching activity is used to estimate power consumption in PrimeTime-2022.03 under typical operating conditions of 25 °C and a core supply voltage of 0.8 V. 

In this section, we evaluate SpikeStream by comparing it against a multi-core SIMD baseline (implementing only the TC, TP, DP, and DB optimization presented in Section \ref{sec:implementation} to assess the speedup and energy savings achieved by the streaming ISA (SA - Section \ref{SA}), and compare SNN inference energy and latency w.r.t. SoA neuromorphic accelerators (Section \ref{subsec:comp_SoA}). 

All the comparisons have been made on a low-latency, single-timestep, S-VGG11 architecture, trained with temporal backpropagation, for image recognition on the CIFAR10 dataset adapted on the work presented by authors \cite{neurips}. The network performs spike encoding using the first layer. To account for the dynamic sparsity of the ifmaps, we conducted our evaluation over a batch of 128 input images, and we reported the computed standard deviations and the average of each considered metric.
\vspace{-1.68em}
\subsection{Performance and memory footprint}
\label{perfmem}
\begin{figure}[t]
    \centering
    \vspace{-0.5em}

    \begin{subfigure}{0.48\textwidth}
        \centering
        \includegraphics[bb=1 2 493 227, width=\linewidth]{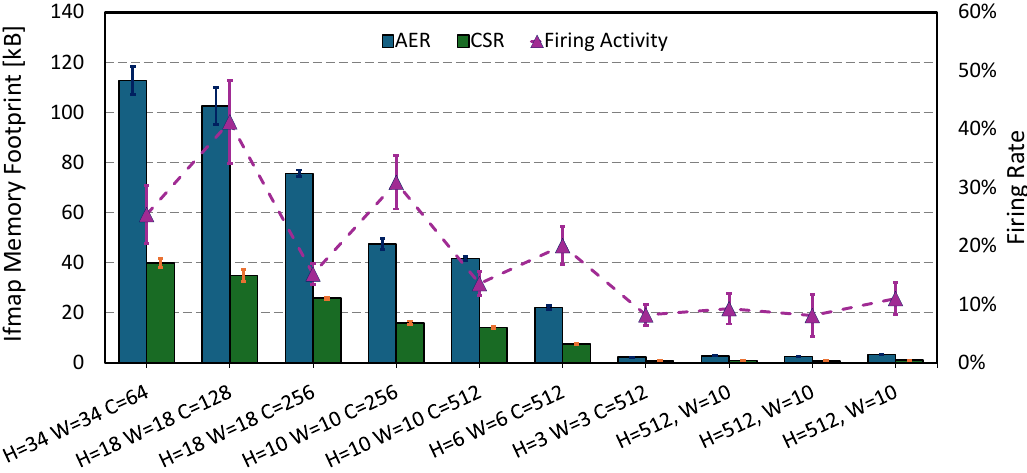}
        \caption{Average firing rate and memory footprint of ifmaps across S-VGG11 layers.}
        \label{fig:frate}
    \end{subfigure}
    \\
    [2ex]

    \begin{subfigure}{0.47\textwidth}
        \centering
        \includegraphics[bb=0 1 505 239, width=\linewidth]{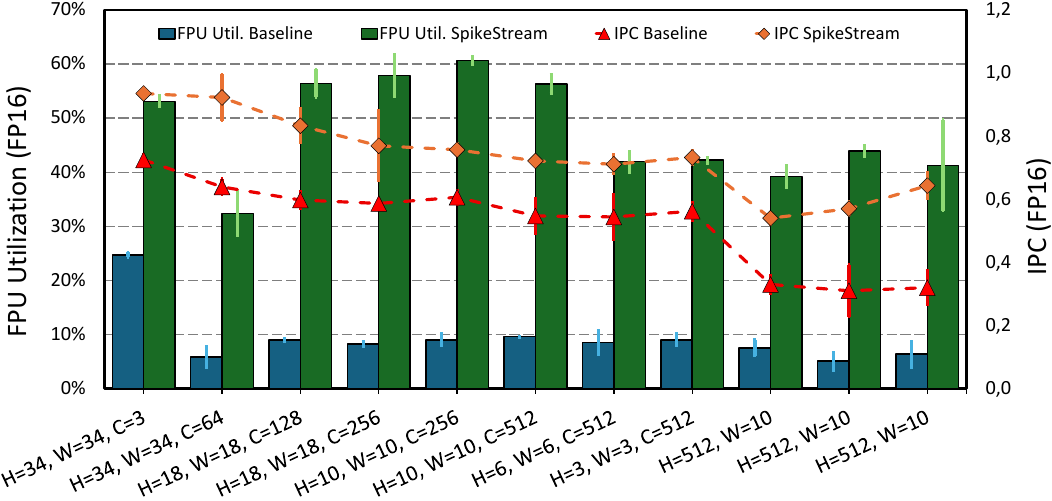}
        \caption{Average FPU utilization and per-core IPC for both code variants in FP16 across S-VGG11 layers.}
        \label{fig:util}
    \end{subfigure}
    \\
    [2ex]

    \captionsetup[subfigure]{justification=centering} 
    
    \begin{subfigure}{0.45\textwidth}
        \centering
        \includegraphics[bb=0 0 495 236, width=\linewidth]{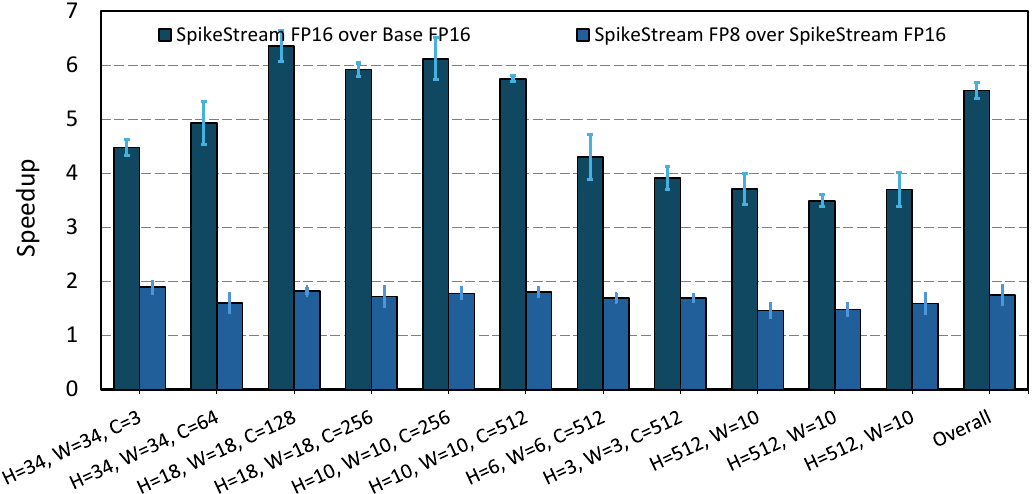}
        \caption{Average speedup across S-VGG11 layers.}
        \label{fig:speedup}
    \end{subfigure}

    \caption{Performance and memory footprint evaluation on S-VGG11, showing average values and standard deviations for a batch size of 128 input frames.}
    \label{fig:perfmem}
    \vspace{-1.5em}
\end{figure}

In this subsection, we evaluate the performance and memory footprint.
Figure \ref{fig:frate} shows the average memory footprint, measured in kB, required to store the ifmaps using both the AER format and our CSR-based format, along with the average firing activity across the various S-VGG11 layers. In both cases, we assume 16-bit values to represent indices and coordinates. The CSR format achieves an average memory footprint reduction of approximately 2.75$\times$ across the different layers of the network when compared to the AER format.

Figure \ref{fig:util} shows the average FPU utilization and IPC distribution across the layers of the network for both SpikeStream and the Baseline with FP16 arithmetic. The first layer, responsible for spike encoding, exhibits the highest FPU utilization for the baseline as convolution is implemented with dense matmul, leading to more regular memory accesses and reduced control computations for the integer core. SpikeStream, using two affine SRs, increases utilization from 24.8\% to 53.1\%.
The second layer exhibits the lowest SpikeStream FPU utilization, primarily due to the reduced number of channels and the sparsity, which further shortens the SpVA stream, not allowing for complete computation overlap and the continuous issue of indirect streams. Indeed, low spike counts result in short stream lengths, leaving the FP pipeline underutilized and resulting in execution time being dominated by the integer pipeline of the snitch cores, which manages the workload-stealing scheduler and the computation of stream base addresses.
In the other convolutional layers, the increasing sparsity is counterbalanced by the larger depth sizes, resulting in an average FPU utilization increase of 6.58$\times$. In the FC layers, the extreme sparsity leads to a marginally lower utilization improvement of SpikeStream.

Figure \ref{fig:speedup} shows the average speedup of SpikeStream when FP16 and FP8 are used w.r.t the baseline in FP16. 
SpikeStream FP16 achieves an average speedup of 5.62$\times$ w.r.t. the baseline for the full S-VGG11 inference, thanks to the SRs and hardware loops utilized in SpikeStream.
Furthermore, in the first two layers of the network, we observe a smaller speedup in SpikeStream FP16 compared to deeper layers. This is primarily due to the limited decoupling between the integer core and FPU, caused by shorter stream lengths in these early layers, which result from their smaller depth sizes and are further reduced by sparsity in the second layer. 
From the third to the sixth network layer, we observe the highest speedup, approaching the ideal of 7$\times$. Here, as with FPU utilization, the sparsity is counterbalanced by the larger depth sizes, allowing full overlap between the integer core and the FPU. The gap to the ideal speedup is mainly due to instruction cache misses, and conflicts in the SPM interconnect resulting from the random access patterns of indirection.
The speedup obtained with SpikeStream in FP8 compared to SpikeStream in FP16 is 1.71$\times$ for the overall network runtime, falling slightly below the ideal of 2$\times$, as two additional iterations are required to unpack the output spikes after the activation function's thresholding.
\subsection{Energy} \label{energy}
\begin{figure}[t]
    \vspace{-0.55em}
    \includegraphics[bb=0 0 489 240, width=0.49\textwidth]{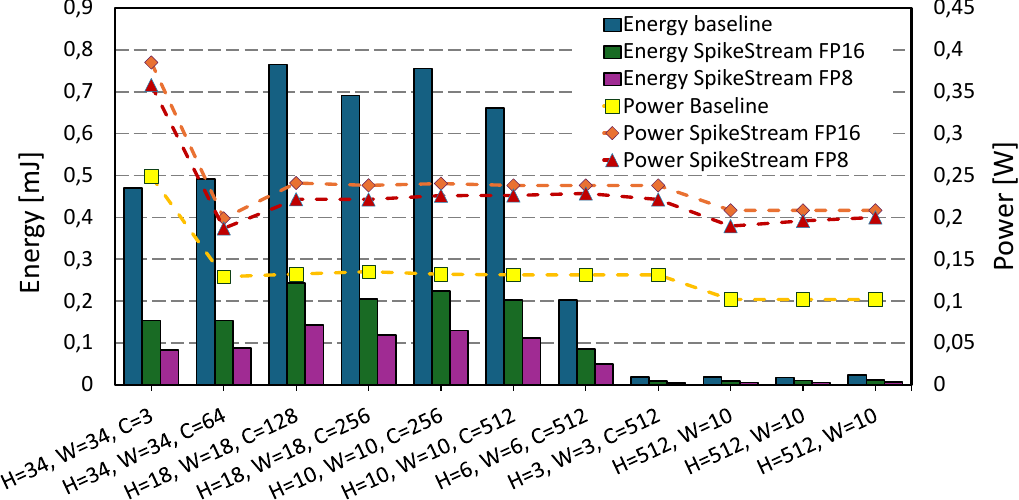}
    \caption{Average energy consumption and power for each layer of the network using both code variants in FP16 and FP8.}
     \label{fig:energy}
\end{figure}
Figure \ref{fig:energy} shows the average energy and power consumption across the various layers of the network for the FP16 baseline, as well as SpikeStream in both FP8 and FP16 formats. The power required in the first layer is higher than in the other layers due to its increased computational intensity, as it involves matmul with multiply-accumulate operations, rather than simpler add operations as in the subsequent layers. Furthermore, the SpikeStream implementations use two affine SRs, instead of one indirect SR used in the following layers, leading to higher power consumption.
Between the second and eighth layers, the power consumption for the evaluated inference kernels (SpikeStream FP8, SpikeStream FP16, Baseline FP16)  remains nearly constant, as the computational kernel is the same, with average values of 0.1319 W, 0.233 W, and 0.219 W for the FP16 baseline, SpikeStream FP16, and SpikeStream FP8, respectively. SpikeStream FP8 also consumes an average of 6.7\% less power than SpikeStream FP16. This reduction is attributed to the way the FPU handles narrower data formats by separating execution units for each format and clock-gating idle slices, reducing power consumption.

Regarding energy consumption, it is primarily concentrated in the convolutional layers of the network, accounting for an average of 82.8\% of the total energy consumption, since the FC layers are smaller and highly sparse. The average efficiency gains in total inference energy consumption are 5.67× for SpikeStream FP8 compared to the baseline, 3.25× for SpikeStream FP16 compared to the baseline, and 1.74× for SpikeStream FP8 compared to SpikeStream FP16.

\subsection{Comparison with SoA neuromorphic accelerators} \label{subsec:comp_SoA}
\begin{figure}[htbp]
    \vspace{-0.3em}
    \centering
    \begin{subfigure}{0.48\columnwidth}
        \centering
        \includegraphics[bb=0 0 446 617, width=\linewidth]{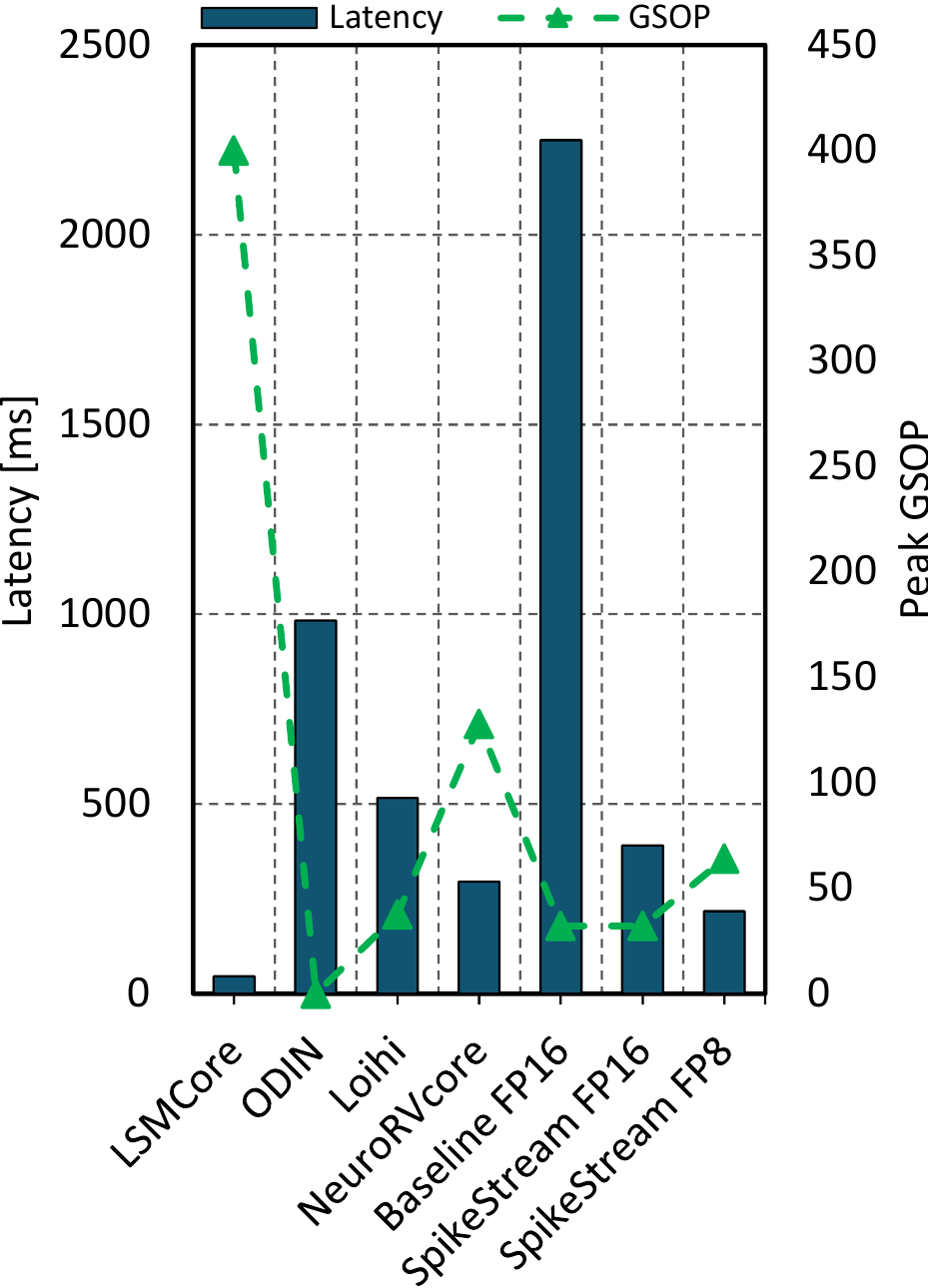}
        \caption{Performance comparison.}
        \label{fig:complat}
    \end{subfigure}
    \hfill
    \begin{subfigure}{0.47\columnwidth}
        \centering
        \includegraphics[bb=1 0 436 621, width=\linewidth]{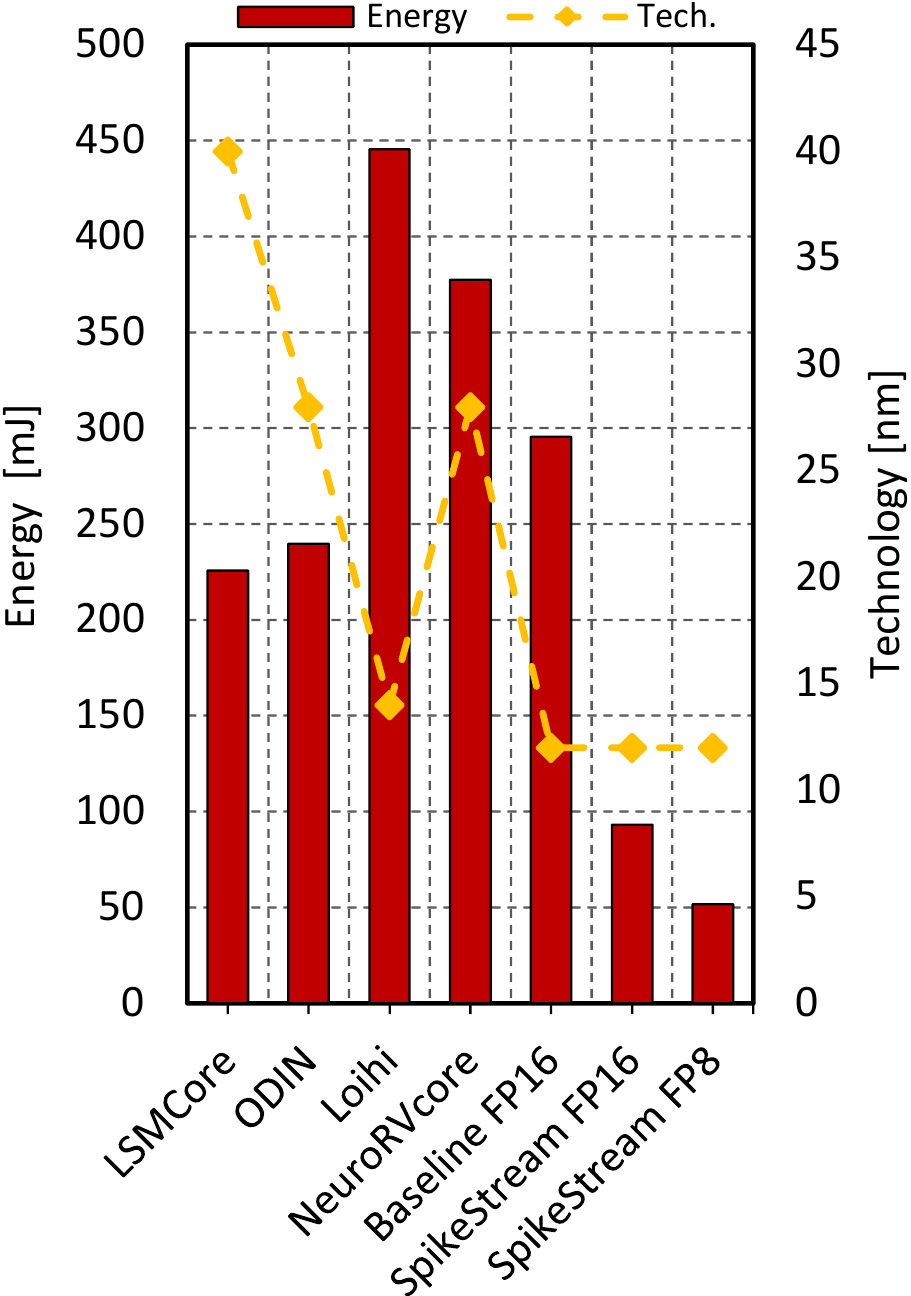}
        \caption{Energy comparison.}
        \label{fig:compen}
    \end{subfigure}
    \vspace{-0.3em}
    \caption{Comparison with neuromorphic processors on the $6^{th}$ layer of S-VGG11 over 500 timesteps.}
    \vspace{-0.68em}
    \label{fig:bothimages}
\end{figure}

In this section, we compare our results with those presented in \cite{neurorv}, which evaluates the energy and runtime for the sixth layer of an S-VGG11 architecture on the CIFAR10 dataset over 500 timesteps on four SoA neuromorphic accelerators: Intel's Loihi (37.5 GSOP, 1-64 bits, 14-nm) \cite{loihi},  ODIN (0.038 GSOP, 4 bits, 28-nm) \cite{odin}, LSMCore (400 GSOP, 4 bits, 40-nm) \cite{lsm}, and NeuroRVcore (128 GSOP, 4 bits, 28-nm) \cite{neurorv}. To ensure a fair comparison, we adapt the neural network accordingly and extend our simulations to match the specified number of timesteps. 
    Figure \ref{fig:complat} shows the performance comparison with neuromorphic accelerators. From the figure, we can notice that among the SoA, LSMCore is the fastest while ODIN is the slowest. This can be easily explained by the peak GSOP of the two architectures, with LSMCore being more than four orders of magnitude more performing. When compared with the proposed architecture, which has 6.25$\times$ (w. FP8 arithmetic) lower peak SOP, the proposed FP16 baseline implementation is the slowest, with a runtime of 2516.72 ms. However, SpikeStream with FP8 arithmetic is the second fastest at 217.14 ms, and only 4.71$\times$ slower than LSMCore, which achieves a latency of only 46.08 ms. This significant result demonstrates the effectiveness of the proposed SR acceleration technique for SNNs.
    This is even more visible when comparing the energy consumption which is reported in Figure \ref{fig:compen}. 
    Our FP16 and FP8 SpikeStream implementations outperform LSMCore (which has the highest energy efficiency among SoA solutions) by consuming 2.37 and 3.46 times less energy, respectively. Despite the different numerical precision which is lower for LSMCore, this can be justified by the more efficient design and technology node of SpikeStream. 
\vspace{-2mm}
\section{Related work}
Among GALS architectures, Intel's Loihi \cite{loihi} is one of the most prominent examples. It features 128 neuromorphic cores and three x86 management cores. Each core contains 1024 spiking neurons arranged in tree structures, mimicking biological elements such as synapses, dendrites, and axons. Communication between cores is handled via an asynchronous Network-on-Chip based on AER format for spike transmission. It occupies a 60 mm² area and is implemented in 14-nm node.

Another GALS architecture that has gained considerable attention is ODIN \cite{odin}. It contains 4 kB of SRAM memory for storing AER spikes and 32 kB of memory for weights. The accelerator's data path contains 64 neurons employing Izhikevich activation functions. The computation is managed by a dedicated event scheduler based on rotating FIFOs. ODIN requires control from an external core connected through an SPI interface and occupies an area of 0.086 mm² in a 28-nm process, achieving an operating frequency of 75 MHz. Dedicated input and output buffers enable it to be integrated into a larger asynchronous mesh.

LSMCore \cite{lsm} implements a DFS solution with 256 input neurons and an array of 1024 LIF neurons. The dynamic sparsity of the ifmaps, stored in a bitmap representation, is handled by zero-skipping weights, while dedicated buffers support ofmaps readout and packing. The design occupies an area of 18.49 mm² in 40-nm technology and operates at a frequency of 400 MHz. Scheduling is handled by a finite-state machine configured by a separate core.

The most versatile solution capable of handling GP workloads is NeuroRVcore \cite{neurorv}. This work extends the RISC-V ISA and the ri5cy core by integrating an accelerator directly into the pipeline. This accelerator features a neuron array to compute the LIF activation function, along with separate neuron-wise and synapse-wise adder trees. It also includes a dedicated vector load/store unit and a vector register file for weights, neuron states and ifmaps. The enhancements introduced by NeuroRVcore result in an area overhead of 149\%, while the arithmetic precision for the weights is fixed to only 4-bit. NeuroRVcore occupies an area of 2.52 mm² and operates at 1 GHz in a 28-nm technology. None of the mentioned solutions are capable of performing online spike encoding, except for neuroRVcore, which can only perform it using the ri5cy pipeline, thus without accelerating it.

\section{Conclusions}
In this paper, we presented the first 
SNN acceleration on a multicore streaming architecture. 
SpikeStream leverages (i) a memory-efficient compression format based on the CSR to represent the sparse ifmaps of SNNs, and (ii) an optimization technique that exploits both affine and indirect SRs combined with decoupling FP hardware loops to accelerate computation and maximize utilization. 
SpikeStream is able to increase utilization to an average of 52.3\% in FP16, achieving a speedup of 7.29× in FP8 and an energy-efficiency gain of 5.68$\times$ in FP8 when compared to a non-streaming implementation. 

Compared to SoA neuromorphic processors, SpikeStream achieves competitive performance in FP8 with a single-layer runtime of 46 ms, compared to 217 ms for LSMCore, which only operates on 4-bit integers. 
SpikeStream outperforms LSMCore with an energy-efficiency gain of 3.46$\times$ in FP8.

Future developments will focus on automatic SpikeStream code generation and enhancing SRs with strided indirect execution to enable higher degrees of computation overlap, further enhancing utilization with extremely sparse ifmaps.


\section{Acknoledgment}
The activity has been funded by Samsung GRO "Enabling Memory Coupled Compute Efficient Acceleration in the Monte Cimone RISC-V cluster" project, the HE EU Graph-Massivizer
(g.a. 101093202) and DECICE (g.a. 101092582) projects.

\bibliographystyle{plain}
\bibliography{ref.bib} 

\begin{thebibliography}{10}

\bibitem{bellec2020}
Guillaume Bellec, Franz Scherr, Anand Subramoney, Elias Hajek, Darjan Salaj, Robert Legenstein, and Wolfgang Maass.
\newblock A solution to the learning dilemma for recurrent networks of spiking neurons.
\newblock {\em Nature communications}, 11(1):3625, 2020.

\bibitem{neurogrid}
Ben~Varkey Benjamin, Peiran Gao, Emmett McQuinn, Swadesh Choudhary, Anand~R. Chandrasekaran, Jean-Marie Bussat, Rodrigo Alvarez-Icaza, John~V. Arthur, Paul~A. Merolla, and Kwabena Boahen.
\newblock Neurogrid: A mixed-analog-digital multichip system for large-scale neural simulations.
\newblock {\em Proceedings of the IEEE}, 102(5):699--716, 2014.

\bibitem{loihi}
Mike Davies, Narayan Srinivasa, Tsung-Han Lin, Gautham Chinya, Yongqiang Cao, Sri~Harsha Choday, Georgios Dimou, Prasad Joshi, Nabil Imam, Shweta Jain, Yuyun Liao, Chit-Kwan Lin, Andrew Lines, Ruokun Liu, Deepak Mathaikutty, Steven McCoy, Arnab Paul, Jonathan Tse, Guruguhanathan Venkataramanan, Yi-Hsin Weng, Andreas Wild, Yoonseok Yang, and Hong Wang.
\newblock Loihi: A neuromorphic manycore processor with on-chip learning.
\newblock {\em IEEE Micro}, 38(1):82--99, 2018.

\bibitem{sne}
Alfio Di~Mauro, Arpan~Suravi Prasad, Zhikai Huang, Matteo Spallanzani, Francesco Conti, and Luca Benini.
\newblock Sne: an energy-proportional digital accelerator for sparse event-based convolutions.
\newblock In {\em 2022 Design, Automation \& Test in Europe Conference \& Exhibition (DATE)}, pages 825--830. IEEE, 2022.

\bibitem{vectstream}
Joao~Mario Domingos, Nuno Neves, Nuno Roma, and Pedro Tomás.
\newblock Unlimited vector extension with data streaming support.
\newblock In {\em 2021 ACM/IEEE 48th Annual International Symposium on Computer Architecture (ISCA)}, pages 209--222, 2021.

\bibitem{odin}
Charlotte Frenkel, Martin Lefebvre, Jean-Didier Legat, and David Bol.
\newblock A 0.086-mm$^2$ 12.7-pj/sop 64k-synapse 256-neuron online-learning digital spiking neuromorphic processor in 28-nm cmos.
\newblock {\em IEEE Transactions on Biomedical Circuits and Systems}, 13(1):145--158, 2019.

\bibitem{cdnn}
Sangyeob Kim and Hoi-Jun Yoo.
\newblock C-dnn v2: Complementary deep-neural-network processor with full-adder/or-based reduction tree and reconfigurable spatial weight reuse.
\newblock {\em IEEE Journal on Emerging and Selected Topics in Circuits and Systems}, 13(4):1026--1039, 2023.

\bibitem{ptb}
Jeong-Jun Lee, Wenrui Zhang, and Peng Li.
\newblock Parallel time batching: Systolic-array acceleration of sparse spiking neural computation.
\newblock In {\em 2022 IEEE International Symposium on High-Performance Computer Architecture (HPCA)}, pages 317--330, 2022.

\bibitem{sobjdetect}
Hong-Han Lien and Tian-Sheuan Chang.
\newblock Sparse compressed spiking neural network accelerator for object detection.
\newblock {\em IEEE Transactions on Circuits and Systems I: Regular Papers}, 69(5):2060--2069, 2022.

\bibitem{dynap}
Saber Moradi, Ning Qiao, Fabio Stefanini, and Giacomo Indiveri.
\newblock A scalable multicore architecture with heterogeneous memory structures for dynamic neuromorphic asynchronous processors (dynaps).
\newblock {\em IEEE Transactions on Biomedical Circuits and Systems}, 12(1):106--122, 2018.

\bibitem{ottati}
Fabrizio Ottati, Chang Gao, Qinyu Chen, Giovanni Brignone, Mario~R. Casu, Jason~K. Eshraghian, and Luciano Lavagno.
\newblock To spike or not to spike: A digital hardware perspective on deep learning acceleration.
\newblock {\em IEEE Journal on Emerging and Selected Topics in Circuits and Systems}, 13(4):1015--1025, 2023.

\bibitem{sssr}
Paul Scheffler, Florian Zaruba, Fabian Schuiki, Torsten Hoefler, and Luca Benini.
\newblock Sparse stream semantic registers: A lightweight isa extension accelerating general sparse linear algebra.
\newblock {\em IEEE Transactions on Parallel and Distributed Systems}, 2023.

\bibitem{ssr}
Fabian Schuiki, Florian Zaruba, Torsten Hoefler, and Luca Benini.
\newblock Stream semantic registers: A lightweight risc-v isa extension achieving full compute utilization in single-issue cores.
\newblock {\em IEEE Transactions on Computers}, 70(2):212--227, 2021.

\bibitem{neuroopp}
Catherine~D. Schuman, Shruti~R. Kulkarni, Maryam Parsa, J.~Parker Mitchell, Prasanna Date, and Bill Kay.
\newblock Opportunities for neuromorphic computing algorithms and applications.
\newblock {\em Nature Computational Science}, 2(1), 1 2022.

\bibitem{lsm}
Lei Wang, Zhijie Yang, Shasha Guo, Lianhua Qu, Xiangyu Zhang, Ziyang Kang, and Weixia Xu.
\newblock Lsmcore: A 69k-synapse/mm2 single-core digital neuromorphic processor for liquid state machine.
\newblock {\em IEEE Transactions on Circuits and Systems I: Regular Papers}, 69(5):1976--1989, 2022.

\bibitem{xiao202}
Mingqing Xiao, Qingyan Meng, Zongpeng Zhang, Di~He, and Zhouchen Lin.
\newblock Online training through time for spiking neural networks.
\newblock {\em Advances in neural information processing systems}, 35:20717--20730, 2022.

\bibitem{neurorv}
Zhijie Yang, Lei Wang, Wei Shi, Yao Wang, Junbo Tie, Feng Wang, Xiang Yu, Linghui Peng, Chao Xiao, Xun Xiao, Yao Yao, Gan Zhou, Xuhu Yu, Rui Gong, Xia Zhao, Yuhua Tang, and Weixia Xu.
\newblock Back to homogeneous computing: A tightly-coupled neuromorphic processor with neuromorphic isa.
\newblock {\em IEEE Transactions on Parallel and Distributed Systems}, 34(11):2910--2927, Nov 2023.

\bibitem{zanatta}
Luca Zanatta, Alfio Di~Mauro, Francesco Barchi, Andrea Bartolini, Luca Benini, and Andrea Acquaviva.
\newblock Directly-trained spiking neural networks for deep reinforcement learning: Energy efficient implementation of event-based obstacle avoidance on a neuromorphic accelerator.
\newblock {\em Neurocomputing}, 562:126885, 2023.

\bibitem{snitch}
Florian Zaruba, Fabian Schuiki, Torsten Hoefler, and Luca Benini.
\newblock Snitch: A tiny pseudo dual-issue processor for area and energy efficient execution of floating-point intensive workloads.
\newblock {\em IEEE Transactions on Computers}, 70(11):1845--1860, 2021.

\bibitem{neurips}
Yaoyu Zhu, Zhaofei Yu, Wei Fang, Xiaodong Xie, Tiejun Huang, and Timoth\'{e}e Masquelier.
\newblock Training spiking neural networks with event-driven backpropagation.
\newblock In {\em Proceedings of the 36th International Conference on Neural Information Processing Systems}, NIPS '22, Red Hook, NY, USA, 2024. Curran Associates Inc.

\end{thebibliography}

\end{document}